\documentclass[final,3p,times]{elsarticle}

\usepackage{amssymb}
\usepackage{amsmath}
\usepackage[normalem]{ulem}
\usepackage{xcolor}
\newcommand{\com}[1]{{\color{black}#1}} 

\journal{Journal of Subatomic Particles and Cosmology}

\begin{document}

\begin{frontmatter}



\title{Charmonium production in p+A collisions at SPS and FAIR energies}

\author[aaa]{Taesoo Song}
\author[bbb,ccc]{Jiaxing Zhao}
\author[ddd,eee]{Joerg Aichelin}
\author[aaa,bbb,ccc]{Elena Bratkovskaya}
\affiliation[aaa]{organization={GSI Helmholtzzentrum f\"{u}r Schwerionenforschung GmbH},
             addressline={Planckstrasse 1},
             city={Darmstadt},
             postcode={64291},
             country={Germany}}

 \affiliation[bbb]{organization={Institute for Theoretical Physics, Johann Wolfgang Goethe Universit\"{a}t},
             addressline={Max-von-Laue-Straße 1},
             city={Frankfurt am Main},
             postcode={60438},
             country={Germany}}

 \affiliation[ccc]{organization={Helmholtz Research Academy Hessen for FAIR (HFHF),GSI Helmholtz Center for Heavy Ion Research. Campus Frankfurt},
             addressline={Max-von-Laue-Straße 14-18},
             city={Frankfurt am Main},
             postcode={60438},
             country={Germany}}

 \affiliation[ddd]{organization={SUBATECH UMR 6457 (IMT Atlantique,  Universit\'{e} de Nantes, IN2P3/CNRS)},
             addressline={4 Rue Alfred Kastler},
             city={Nantes},
             postcode={F-44307},
             country={France}}

 \affiliation[eee]{organization={Frankfurt Institute for Advanced Studies},
             addressline={Ruth-Moufang-Strasse 1},
             city={Frankfurt am Main},
             postcode={60438},
             country={Germany}}

\begin{abstract}

We employ the Parton–Hadron–String Dynamics (PHSD)
transport approach to investigate the influence of baryon-rich matter on
charmonium production and dissociation. The Remler coalescence formalism
is implemented to dynamically model charmonium formation from
charm–anticharm pairs. As a validation step, the formalism is first
benchmarked against experimental data from elementary p+p collisions and
then extended to p+A systems to extract the effective nuclear absorption
cross section of charmonium. This extracted cross section can
subsequently be applied in A+A collisions to quantify medium-induced effects.
Our results demonstrate that the Remler formalism provides a
quantitatively consistent description of charmonium production in p+p and p+A collisions at SPS
energies. 
The approach is then
extrapolated to GSI/FAIR energies, where predictions for charmonium
yields and survival probabilities are presented. These findings
highlight the relevance of the Remler formalism as a dynamical framework
for studying heavy-quark bound-state formation in 
baryon-rich matter and offer theoretical guidance for future
experimental programs at SPS, FAIR and NICA aimed at mapping the QCD phase
structure.
\end{abstract}



\begin{keyword}
Quark Gluon Plasma, Relativistic Heavy Ion Collision, Strangeness and Heavy Flavor


\end{keyword}

\end{frontmatter}



\section{Introduction}
\label{Intro}

Charmonium has long been recognized
as a sensitive probe of the properties of the hot and dense strongly
interacting matter created in relativistic heavy-ion collisions, following 
the pioneering work of Matsui and Satz~\cite{Matsui:1986dk}.
At finite baryon chemical potential ($\mu_B$), lattice QCD predicts a
smooth crossover between hadronic and partonic matter,
whereas at larger $\mu_B$, beyond the conjectured critical end point,
the transition becomes first order. Studying charmonium
behavior in baryon-rich systems—such as those accessible at SPS and the
upcoming GSI/FAIR facilities—therefore provides a unique opportunity to
explore the QCD phase diagram in regions of high net-baryon density.
\com{As a first step, p+A collisions at these energies serve as a clean reference baseline that is essential for disentangling cold/dense nuclear–matter effects from those due to the QGP.}

\section{Remler formalism in PHSD}
\label{Remler}

Our study is based on Remler formalism incorporated dynamically in the PHSD \cite{Cassing:2008sv,Moreau:2019vhw} which is microscopic off-shell transport approach for the dynamical description of  non-equilibrium evolution of strongly interacting matter. PHSD consistently propagates partonic and hadronic degrees-of-freedom and their interactions based on Kadanoff–Baym theory. The non-perturbative QGP is described within the Dynamical QuasiParticle Model (DQPM) \cite{Moreau:2019vhw} in terms of $(T,\mu_B)$-dependent massive quarks and gluons with finite spectral widths, formulated in a two-particle irreducible (2PI) propagator representation.

\begin{figure}[t]
\centerline{
\includegraphics[width=8 cm]{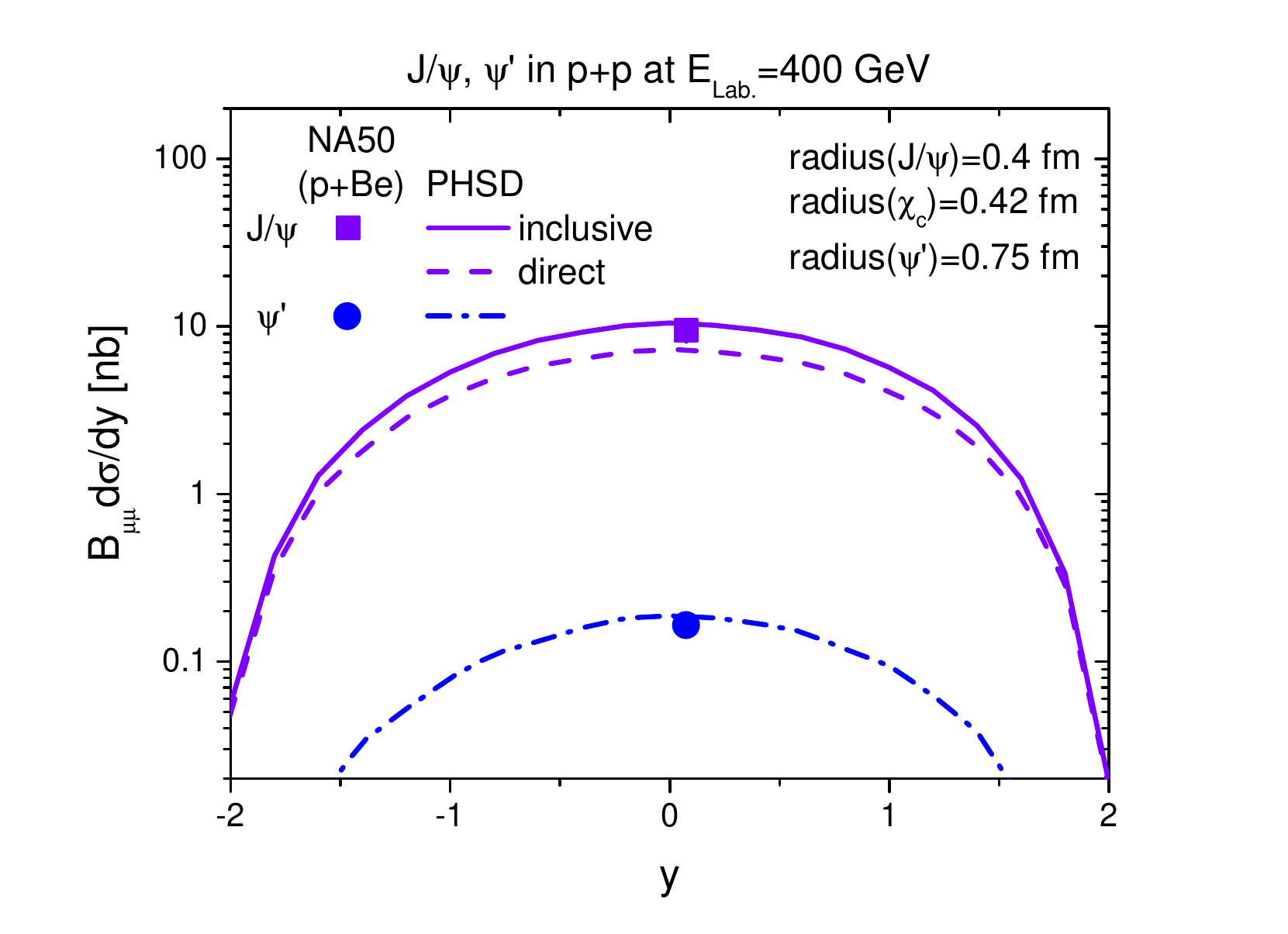}
\includegraphics[width=8 cm]{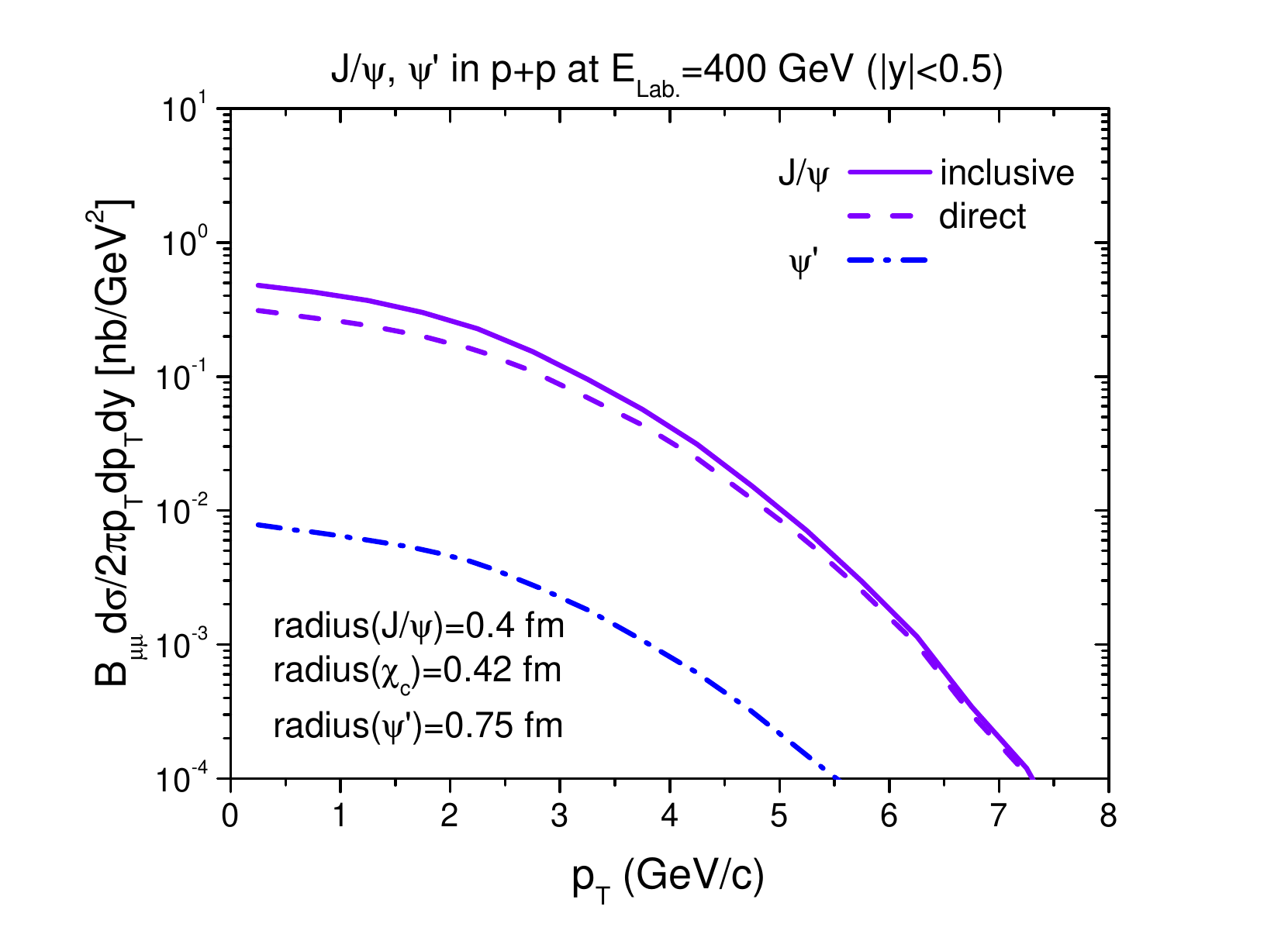}}
\caption{Differential cross sections for charmonium production, multiplied by the branching ratio to dimuon decay, as a function of rapidity and transverse momentum in p+p collisions at $\sqrt{s_{NN}}=$ 27 GeV ($E_{kin}=$ 400 GeV). The experimental data are taken from the NA50~\cite{NA50:2006rdp}}\label{sigma-pp}
\end{figure}

\begin{figure}[t]
\centerline{
\includegraphics[width=8 cm]{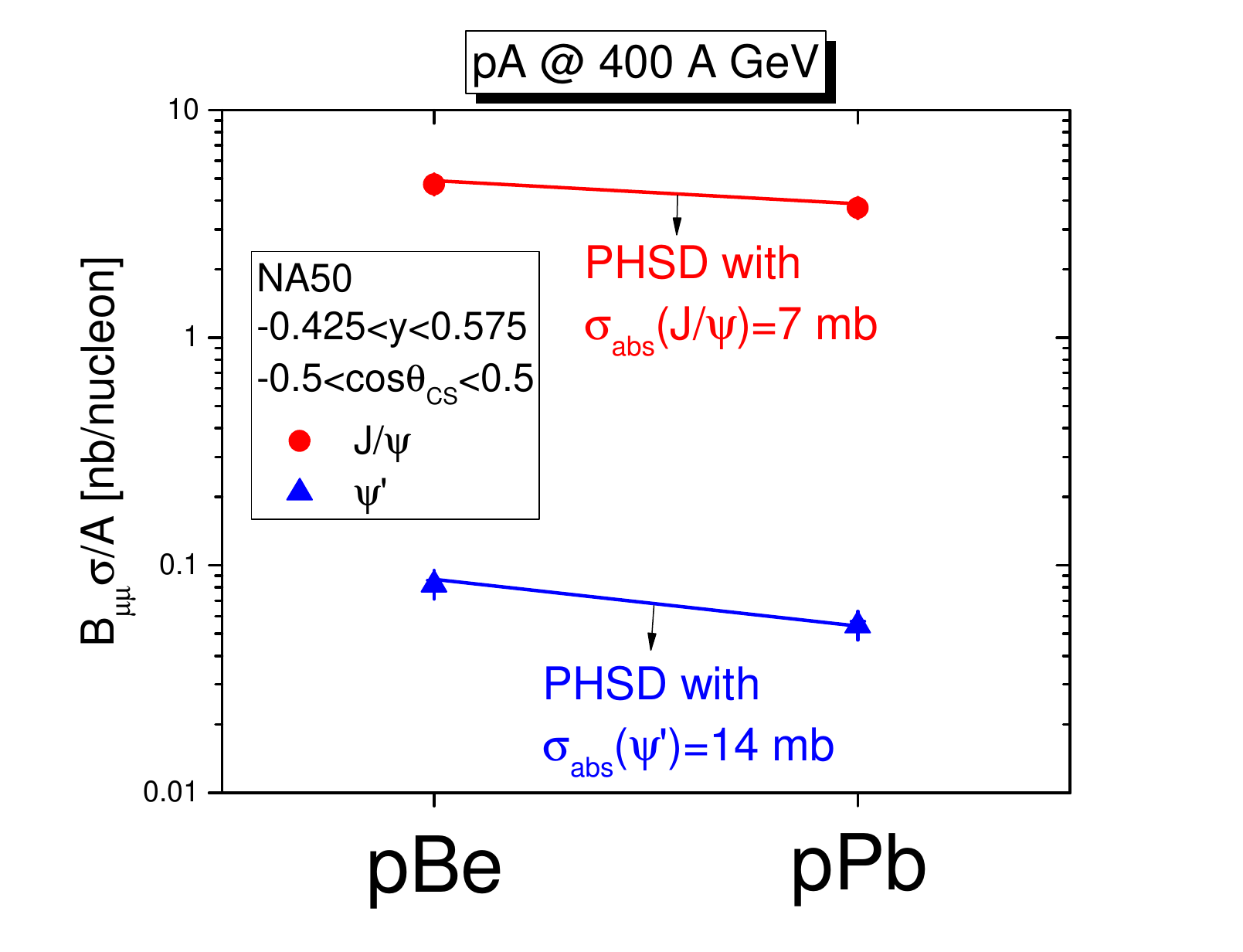}
\includegraphics[width=8 cm]{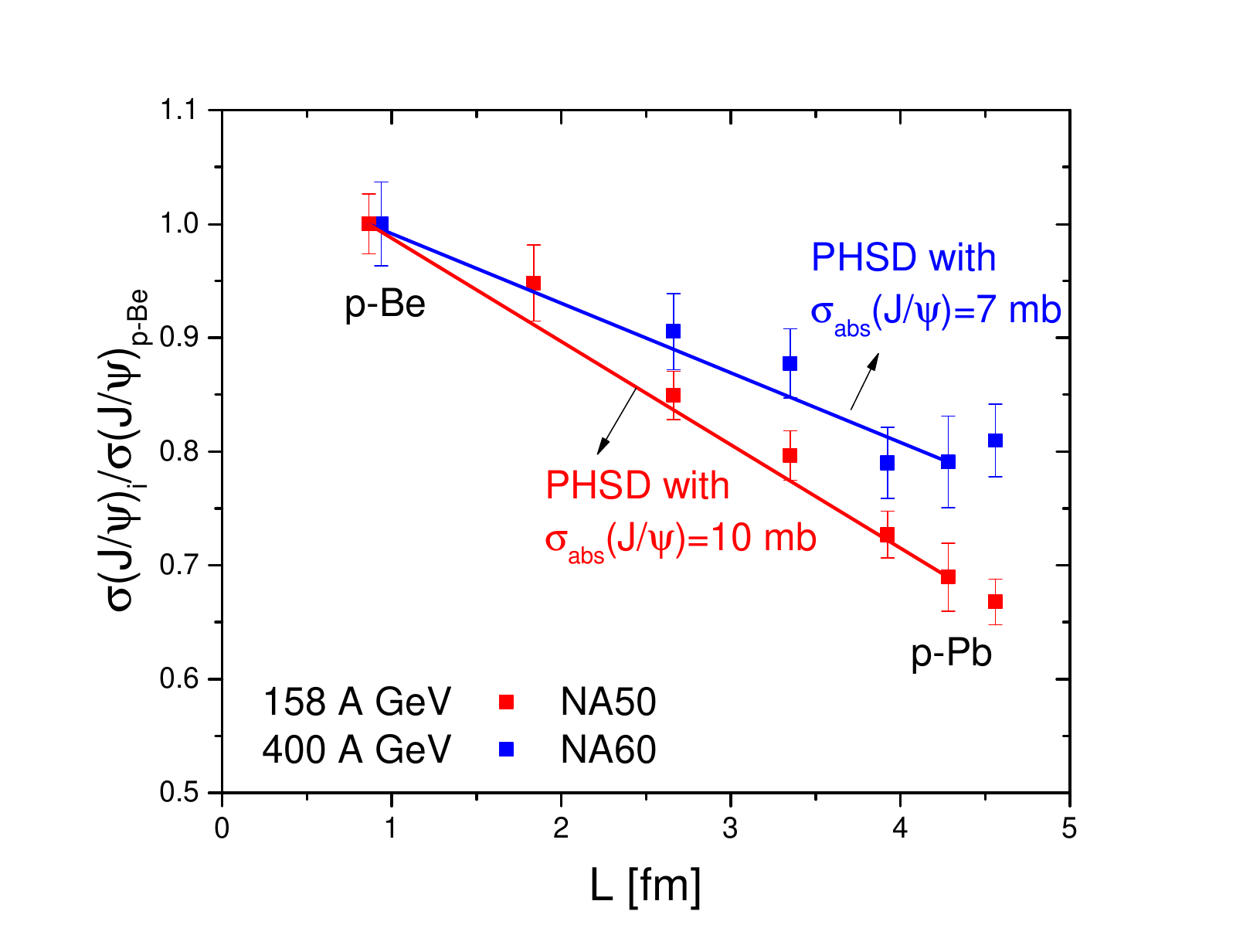}}
\caption{(Left) production cross sections of $J/\psi$ and $\psi'$, multiplied by their branching ratios to dimuons and divided by the target mass number ($A$), in p+Be and p+Pb collisions at $E_{kin}$=400 GeV from the NA50 Collaboration~\cite{NA50:2006rdp}, and (right) the ratio of $J/\psi$ production cross sections in p+A collisions to those in p+Be collisions as a function of nuclear path length at $E_{kin}$=400 and 158 GeV~\cite{NA60:2010wey}.}\label{fig2}
\end{figure}

\begin{figure}[t]
\centering{
\includegraphics[width=8 cm]{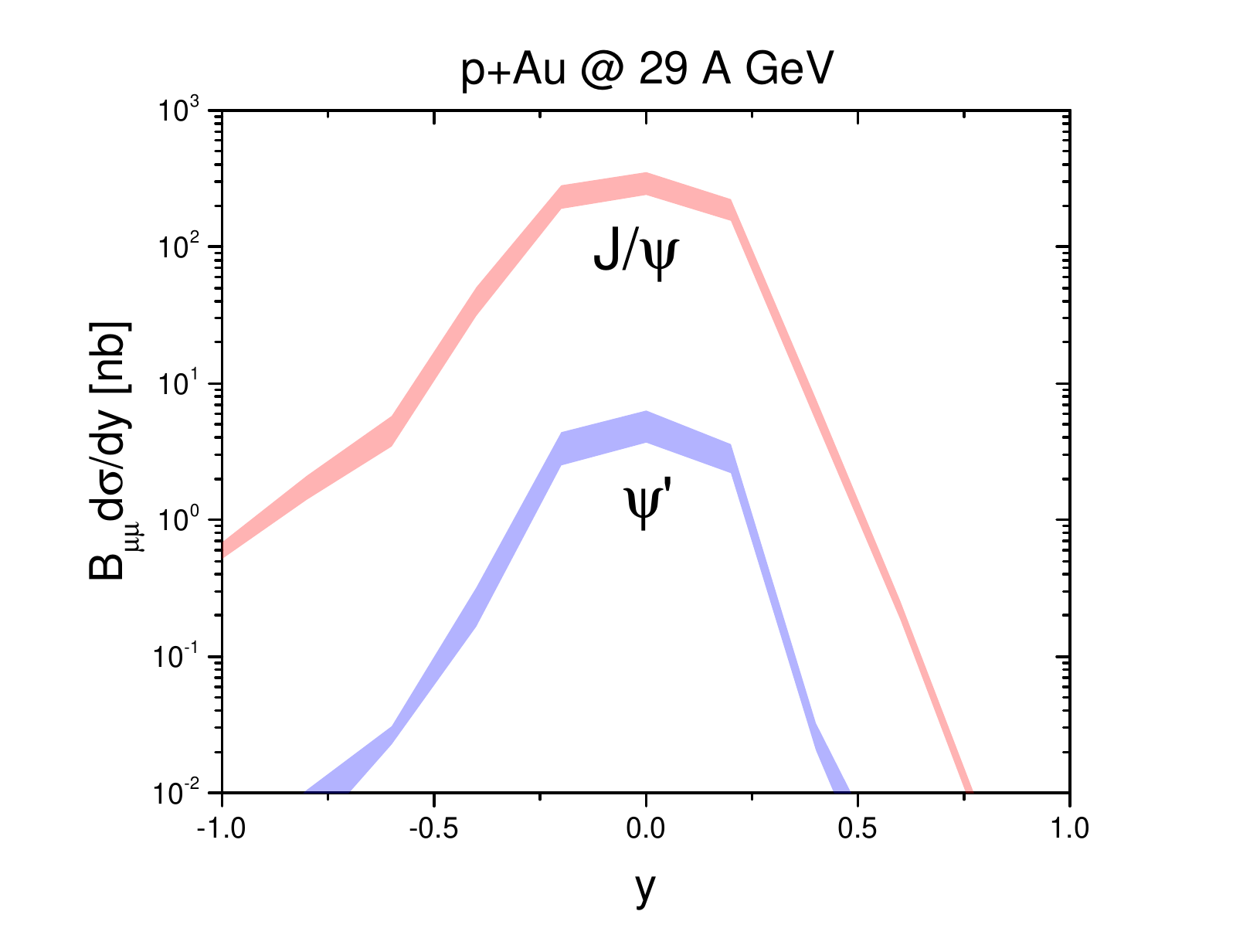}
\includegraphics[width=8 cm]{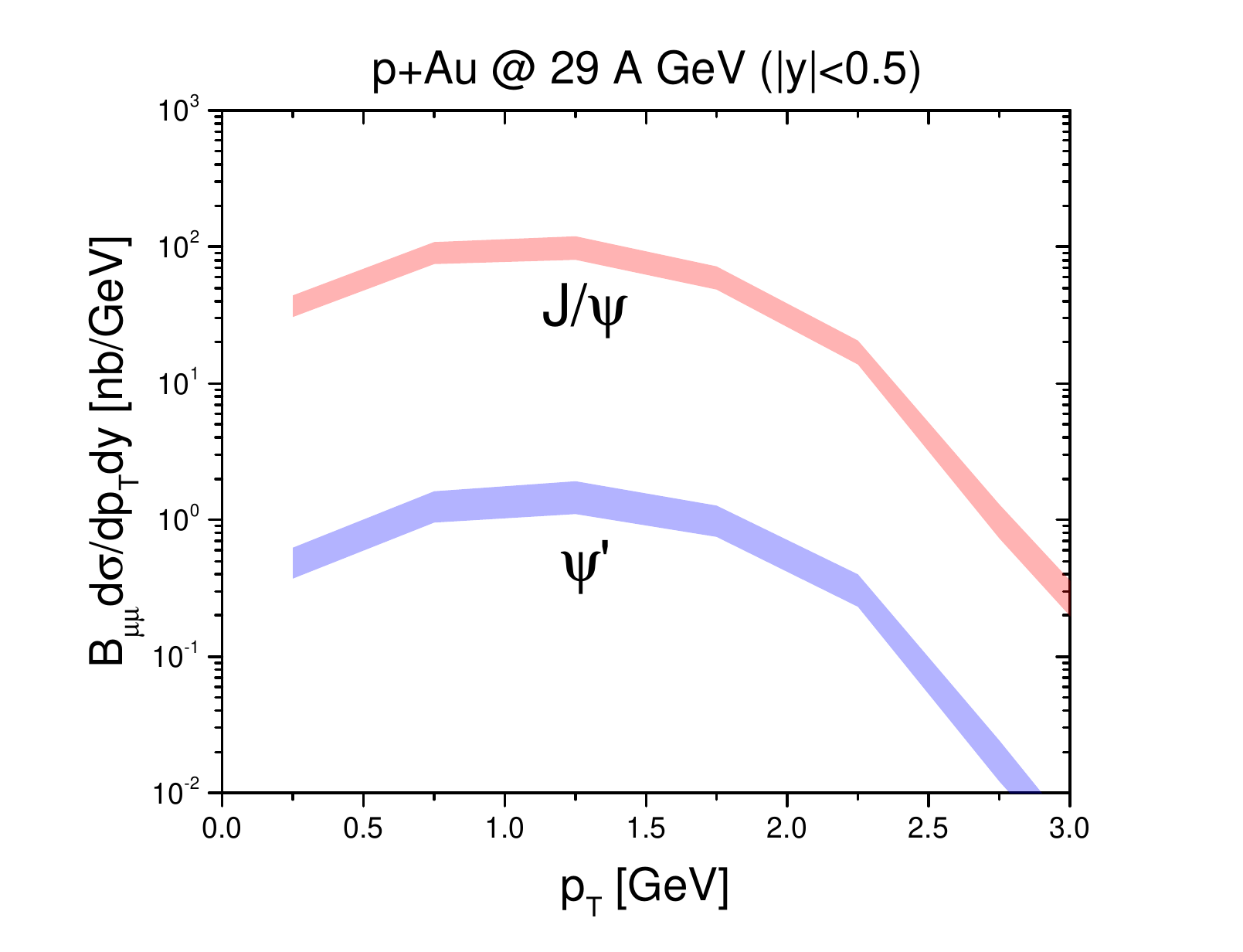}}
\caption{Production cross sections of $J/\psi$ and $\psi^\prime$, multiplied by their respective branching ratios to dimuons, as functions of rapidity and transverse momentum in p+Au collisions at E=29 GeV. Absorption cross sections of 10-20 mb are assumed for $J/\psi$ and $\chi_c$, and 20-40 mb for $\psi^\prime$. \com{The upper and lower boundaries of the bands correspond to the smaller and larger absorption cross sections, respectively.}}\label{fair}
\end{figure}

In the Remler formalism~\cite{Remler:1975re,Remler:1975fm}, the number of quarkonia is given by 
\begin{equation}
    N_\Phi(t) = N_\Phi(0)+\int_0^t \Gamma(t^\prime)dt^\prime,
\label{probability}    
\end{equation}
where $t$ denotes the hadronization time of heavy quark, and the rate $\Gamma(t)$ is expressed as
\begin{eqnarray}
\com{\Gamma(t)
= \sum_{i,j}\sum_{\nu_{i},\nu_{j}}\frac{1}{(2\pi)^{3N}}  \int d^3r_1d^3p_1 ... d^3r_N d^3p_N W_0(\vec{r},\vec{p})
\bigg\{\delta(t-t^\nu_{i})
-\delta(t-t^{\nu-1}_{i})+\delta(t-t^\nu_{j})
-\delta(t-t^{\nu-1}_{j})\bigg\}W^{(N)}(t+\varepsilon),}
\label{new}
\end{eqnarray}
where $N$ is the total number of heavy quarks and heavy antiquarks, indexed by $i$ and $j$, respectively.
Here, $t^\nu_{i(j)}$ denotes the time of $\nu$th scattering of the heavy quark $i$ (or heavy antiquark $j$) in the QGP. 
$W_0$ is the Wigner function of the quarkonium \com{with $\vec{r}=\vec{r}_i-\vec{r}_j$, $\vec{p}=\vec{p}_i-\vec{p}_j$,} whereas $W^{(N)}(t)$ is the density matrix of the $N$ heavy quarks and heavy antiquarks, which is approximated by their classical phase-space density distribution~\cite{Song:2026pld}.
The parameter $\varepsilon$ in Eq.~(\ref{new}) indicates that the Wigner density
is updated after the momentum change of a heavy (anti)quark resulting from a collision in the QGP. 
Considering the thermal interaction between quarkonium and the medium, the first and third terms in the curly bracket of Eq.~(\ref{new}) can be interpreted as quarkonium regeneration, while the second and fourth terms correspond to thermal dissociation.

From Eqs.~(\ref{probability}) and (\ref{new}), the initial Wigner projection in the absence of subsequent scattering, is given by
\begin{eqnarray}
N_\Phi(t_0)
= \sum_{i,j}\frac{1}{(2\pi)^{3N}}  \int d^3r_1d^3p_1 ... d^3r_N d^3p_N W_0(\vec{r}_i-\vec{r}_j,\vec{p}_i-\vec{p}_j)
W^{(N)}(t_0),
\label{begin}
\end{eqnarray}
where $t_0$ is the time of the initial Wigner projection. 
Assuming the wavefunction of a 3-dimensional harmonic oscillator,
the Wigner densities of a $S-$state and a $P-$state are given by
\begin{eqnarray}
W_0^{\rm S}({\bf r, p})=8\frac{D}{d_1 d_2}\exp\bigg[-\frac{r^2}{\sigma^2}-\sigma^2p^2\bigg],~~~~~~~
W_0^{\rm P}({\bf r, p})=\frac{16}{3}\frac{D}{d_1 d_2}\bigg(\frac{r^2}{\sigma^2}-\frac{3}{2}+\sigma^2p^2\bigg)\exp\bigg[-\frac{r^2}{\sigma^2}-\sigma^2p^2\bigg].
\label{wigner2}
\end{eqnarray}
The parameter $\sigma$ is constrained by $\sigma^2=2/3\langle r^2\rangle$ for an $S-$state and $\sigma^2=2/5\langle r^2\rangle$ for a $P-$state with $\sqrt{\langle r^2\rangle}$ being the root-mean-square (rms) distance.
The factors $D$, $d_1$, and $d_2$ denote the color-spin degeneracies of charmonium, the charm quark, and the charm antiquark, respectively.
The Wigner function for a $P-$state can become locally negative. 
\com{To avoid the issue of negative probabilities, these negative values are set to zero, which consequently enhances $\chi_c$ production.}
In addition, the Wigner function for a $2S$ state is simplified into the form of $1S$.
The momenta of the charm and anticharm quarks are given by the PYTHIA event generator, and their spatial separation is sampled from a Gaussian distribution whose mean-square radius is inversely proportional to the charm-quark mass~\cite{Song:2026pld}.

\section{Results}

Fig.~\ref{sigma-pp} shows the differential cross section for charmonium production as functions of rapidity and transverse momentum in p+p collisions  at $E_{kin}=$ 400 GeV. The calculations assume radii, defined as ($\sqrt{\langle r^2\rangle}/2$), of 0.4, 0.42, and 0.75 fm for $J/\psi$, $\chi_c$ and $\psi^\prime$, respectively~\cite{Song:2026pld}.
Since no experimental data are available for p+p collisions at SPS energies, the PHSD results are compared with p+Be collision data from the NA50 Collaboration~\cite{NA50:2006rdp}, normalized by the mass number of Beryllium. We note that the PHSD results are divided by a factor of two to account for the experimental acceptance condition $-0.5<\cos\theta_{CS}<0.5$, where $\theta_{CS}$ is the polar angle in the Collins-Soper frame.

The Remler formalism for charmonium production is applied only in the partonic phase. 
Once the partons hadronize, the charmonia produced through the Remler formalism undergo hadronic interactions, which can be categorized into nuclear absorption and comover effects, the latter referring to interactions with produced mesons. 
Nuclear absorption corresponds to the inelastic scattering of charmonium off baryons and is the dominant suppression mechanism for charmonium in p+A collisions. 
In principle, the nuclear absorption cross section depends on the scattering energy. For simplicity, however, it is assumed to be constant at a given p+A collision energy and is determined from experimental data~\cite{NA50:2006rdp,NA60:2010wey}.

The left panel of Fig.~\ref{fig2} presents the PHSD results for the production cross sections of $J/\psi$ and $\psi'$, multiplied by their branching ratios to dimuons and divided by the target mass number \com{($A$)}, in p+Be and p+Pb collisions at $E_{kin}$=400 A GeV~\cite{Song:2026pld}. 
Since Be is a very light nucleus $(A=9)$, nuclear absorption has only a small impact on charmonium production, whereas much stronger effects are expected for a Pb target.
The figure shows that the PHSD results are consistent with the NA50 data for p+Be collisions, as already demonstrated in Fig.~\ref{sigma-pp}.
Furthermore, nuclear absorption cross sections of 7 mb for direct $J/\psi$ and 14 mb for direct $\psi^\prime$ reproduce the measured charmonium production cross sections in p+Pb collisions.
Since no experimental data are available for $\chi_c$, we assume that its absorption cross section is the same as that of $J/\psi$. 
The right panel shows the ratio of the $J/\psi$ production cross section per nucleon in p+A collisions at $E_{kin}$= 400 and 158 GeV to that in p+Be collisions as a function of the average path length of nuclear matter traversed by the produced $J/\psi$.
In the figure, the ratio is evaluated only for p+Pb collisions ($L\approx 4.3$), and straight lines are drawn to facilitate a qualitative comparison with experimental data from the NA50 and NA60 Collaborations~\cite{NA60:2010wey}.
We find that an absorption cross section of 7~mb for $J/\psi$ is required at $E_{kin}$= 400 GeV, whereas the value increases to 10 mb at $E_{kin}$= 158 GeV. This indicates that a larger absorption cross section is required at lower collision energies to reproduce the experimental data.
For $\psi^\prime$, we assume that the absorption cross section at $E_{kin}$= 158 GeV is twice that of $J/\psi$, as is also the case at $E_{kin}$= 400 GeV.

Finally, turning to the FAIR energies, Fig.~\ref{fair} shows the production cross sections of $J/\psi$ and $\psi^\prime$, multiplied by their branching ratios to dimuons, as functions of rapidity and transverse momentum at midrapidity ($|y|<0.5$) in p+Au collisions at E=29 GeV.
The calculations assume absorption cross sections of 10-20 mb for $J/\psi$ and $\chi_c$ and 20-40 mb for $\psi^\prime$.
\com{Comover effects from the produced mesons are also included, but they play only a minor role in p+A collisions.}

\section*{Acknowledgements}
We acknowledge support by the Deutsche Forschungsgemeinschaft (DFG, German Research Foundation) through the grant CRC-TR 211 'Strong-interaction matter under extreme conditions' - Project number 315477589 - TRR 211. 
The computational resources have been provided by the LOEWE-Center for Scientific Computing and the "Green Cube" at GSI, Darmstadt and by the Center for Scientific Computing (CSC) of the Goethe University.
The publication is funded by the Open Access Publishing Fund of GSI Helmholtzzentrum fuer Schwerionenforschung.

\bibliographystyle{elsarticle-num}
\bibliography{sqm2026_song}

@article{Matsui:1986dk,
    author = "Matsui, T. and Satz, H.",
    title = "{$J/\psi$ Suppression by Quark-Gluon Plasma Formation}",
    reportNumber = "BNL-38344",
    doi = "10.1016/0370-2693(86)91404-8",
    journal = "Phys. Lett. B",
    volume = "178",
    pages = "416--422",
    year = "1986"
}

@article{Remler:1975re,
    author = "Remler, E. A. and Sathe, A. P.",
    title = "{Quasi-Classical Scattering Theory and Bound State Production Processes}",
    doi = "10.1016/0003-4916(75)90223-7",
    journal = "Annals Phys.",
    volume = "91",
    pages = "295--324",
    year = "1975"
}

@article{Remler:1975fm,
    author = "Remler, E. A.",
    title = "{Use of the Wigner Representation in Scattering Problems}",
    doi = "10.1016/0003-4916(75)90065-2",
    journal = "Annals Phys.",
    volume = "95",
    pages = "455--495",
    year = "1975"
}

@article{Song:2026pld,
    author = "Song, Taesoo and Zhao, Jiaxing and Aichelin, Joerg and Bratkovskaya, Elena",
    title = "{Charmonium production at SPS and FAIR energies}",
    eprint = "2605.29479",
    archivePrefix = "arXiv",
    primaryClass = "hep-ph",
    month = "5",
    year = "2026"
}

@article{NA50:2006rdp,
    author = "Alessandro, B. and others",
    collaboration = "NA50",
    title = "{J/psi and psi-prime production and their normal nuclear absorption in proton-nucleus collisions at 400-GeV}",
    eprint = "nucl-ex/0612012",
    archivePrefix = "arXiv",
    reportNumber = "CERN-PH-EP-2006-018",
    doi = "10.1140/epjc/s10052-006-0079-4",
    journal = "Eur. Phys. J. C",
    volume = "48",
    pages = "329",
    year = "2006"
}

@article{NA60:2010wey,
    author = "Arnaldi, R and others",
    collaboration = "NA60",
    title = "{J/psi production in proton-nucleus collisions at 158 and 400 GeV}",
    eprint = "1004.5523",
    archivePrefix = "arXiv",
    primaryClass = "nucl-ex",
    doi = "10.1016/j.physletb.2011.11.042",
    journal = "Phys. Lett. B",
    volume = "706",
    pages = "263--267",
    year = "2012"
}

@article{Cassing:2008sv,
    author = "Cassing, W. and Bratkovskaya, E. L.",
    title = "{Parton transport and hadronization from the dynamical quasiparticle point of view}",
    eprint = "0808.0022",
    archivePrefix = "arXiv",
    primaryClass = "hep-ph",
    doi = "10.1103/PhysRevC.78.034919",
    journal = "Phys. Rev. C",
    volume = "78",
    pages = "034919",
    year = "2008"
}

@article{Moreau:2019vhw,
    author = "Moreau, Pierre and Soloveva, Olga and Oliva, Lucia and Song, Taesoo and Cassing, Wolfgang and Bratkovskaya, Elena",
    title = "{Exploring the partonic phase at finite chemical potential within an extended off-shell transport approach}",
    eprint = "1903.10257",
    archivePrefix = "arXiv",
    primaryClass = "nucl-th",
    doi = "10.1103/PhysRevC.100.014911",
    journal = "Phys. Rev. C",
    volume = "100",
    number = "1",
    pages = "014911",
    year = "2019"
}



\end{document}